\documentclass[aps,prd,preprint,superscriptaddress,tightenlines,nofootinbib]{revtex4}

\usepackage{graphicx}
\usepackage{dcolumn}
\usepackage{bm}

\begin{document}

\newcommand{\gga}{\gamma\gamma}
\newcommand{\piz}{\pi^0}
\newcommand{\tpi}{\pi^+\pi^-\piz}
\newcommand{\tpz}{3\piz}
\newcommand{\eeg}{e^+e^-\gamma}
\newcommand{\ppg}{\pi^+\pi^-\gamma}
\newcommand{\ppee}{\pi^+\pi^- e^+e^- }
\newcommand{\ppgg}{\pi^+\pi^-\gamma\gamma}
\newcommand{\invis}{\mbox{invisible}}

\newcommand{\invpb}{\,\mathrm{pb}^{-1}}
\newcommand{\gev}{\,\mathrm{GeV}}
\newcommand{\mev}{\,\mathrm{MeV}}
\newcommand{\mrad}{\,\mathrm{mrad}}

\newcommand{\chijv}{\chi_{J/\psi, \mathrm{v}}^2/\mathrm{d.o.f.}}
\newcommand{\chijm}{\chi_{J/\psi, \mathrm{m}}^2/\mathrm{d.o.f.}}
\newcommand{\chipv}{\chi_{\psi(2S), \mathrm{v}}^2/\mathrm{d.o.f.}}
\newcommand{\chipm}{\chi_{\psi(2S), \mathrm{m}}^2/\mathrm{d.o.f.}}

\newcommand{\etal}{{\it et al.}}

\preprint{CLNS 07/2002}       
\preprint{CLEO 07-10}        

\title{Measurement of prominent $\eta$ decay branching fractions}

\author{A.~Lopez}
\author{S.~Mehrabyan}
\author{H.~Mendez}
\author{J.~Ramirez}
\affiliation{University of Puerto Rico, Mayaguez, Puerto Rico 00681}
\author{J.~Y.~Ge}         
\author{D.~H.~Miller}
\author{B.~Sanghi}
\author{I.~P.~J.~Shipsey}
\author{B.~Xin}
\affiliation{Purdue University, West Lafayette, Indiana 47907, USA}
\author{G.~S.~Adams}
\author{M.~Anderson}
\author{J.~P.~Cummings}
\author{I.~Danko}
\author{D.~Hu}
\author{B.~Moziak}
\author{J.~Napolitano}
\affiliation{Rensselaer Polytechnic Institute, Troy, New York 12180, USA}
% ...
\collaboration{CLEO Collaboration} 
\noaffiliation

\date{July 9, 2007}

\begin{abstract} 
The decay $\psi(2S) \to\eta J/\psi$ is used to measure,
for the first time, all prominent $\eta$-meson branching
fractions with the same experiment in the same dataset,
thereby providing a consistent treatment of systematics
across branching fractions. 
We present results for $\eta$ decays to $\gga$, $\tpi$,
$\tpz$, $\ppg$, and $\eeg$, accounting for $99.9\%$ of all
$\eta$ decays. 
The precisions for several of the branching fractions and 
their ratios 
are improved. Two channels, $\ppg$ and $\eeg$, show results
that differ 
at the level of three standard deviations
from those previously determined.
\end{abstract}

\maketitle

The $\eta$ meson was discovered almost half a century ago~\cite{Pevsner:1961pa}. 
It is the second-lightest meson, considered to consist of $u$, $d$,
and $s$ quarks, and studying its decays into pions,
electrons, and photons gives insight into different aspects of
non-perturbative QCD and electromagnetic phenomena.
Measurements of the $\eta$ decay properties come from many
different experiments, and almost all exclusive branching fraction
determinations are made relative to other $\eta$ decays. The
Particle Data Group (PDG)~\cite{Yao:2006px} uses 43 such measurements
in a fit to determine the branching fractions
to $\gga$, $\tpz$, $\tpi$, $\ppg$, $\piz\gamma\gamma$, $\eeg$, 
$\mu^+\mu^-\gamma$, 
and $\ppee$, as well as the total width.

The analysis presented here studies $\eta$ decays in the
reaction $e^+ e^- \to \psi(2S) \to \eta J/\psi$ with
$\eta \to \gga$, $\tpz$, $\tpi$, $\ppg$ and $\eeg$. 
We identify the $J/\psi$ through its decays to $e^+e^-$ and $\mu^+\mu^-$. 
The choice of modes addresses the known branching fractions of 
${\cal O}(0.1\%)$ and larger, and covers 99.88\% of
the $\eta$ decay modes when using the branching fractions 
from Ref.~\cite{Yao:2006px}. 
The strength of this analysis lies in the simultaneous
and similar treatment of charged and neutral $\eta$
decay products, cross-feed of different modes into each other, 
and, with the same
analysis procedure, estimates of backgrounds from other
$X J/\psi$ sources.

The CLEO-c detector is described in detail elsewhere~\cite{CESRCLEO}.
Its features exploited here are the 93\%
solid angle coverage of precision
charged particle tracking and an electromagnetic calorimeter
consisting of 7784 CsI(Tl) crystals,
the barrel portion of which has a vertex-pointing geometry.
The barrel calorimeter and two open-cell 
drift chambers are concentric with
the colliding beams and embedded inside a 1~T
axial magnetic field provided by a superconducting
solenoid. The small inner chamber has six cylindrical stereo layers
(drift cells canted at an angle to the chamber axis),
and the outer, larger chamber has 47 layers,
the inner 16 of which are axial
and the outer 31 stereo. 
(About 5\% of the data used here were
acquired in the earlier CLEO~III detector configuration,
which differed from CLEO-c primarily by
having a four-layer silicon strip
vertex detector in place of the inner tracking chamber.)
The tracking system enables momentum measurements
for particles with momentum
transverse to the beam exceeding 50~MeV/$c$,
and achieves resolution $\sigma_p/p\simeq$0.6\% at $p$=1~GeV/$c$.
The barrel calorimeter reliably measures photon shower energies down
to $E_\gamma$=30~MeV and has a resolution of
$\sigma_E/E\simeq$5\% at 100~MeV and 2.2\% at 1~GeV.

The data sample comprises about 27M $\psi(2S)$ decays,
corresponding to about 0.1M $\eta$ decays produced with
a $J/\psi \to \ell^+\ell^-$ tag.

We determine the detection efficiency and background levels
with Monte Carlo (MC) samples that were generated using the
{\sc EvtGen} event generator~\cite{evtgen}
and a {\sc GEANT}-based~\cite{geant} 
detector simulation. We model $\eta \to \gga$ and $\tpz$
according to phase space. The mode $\ppg$ is
simulated as mediated by a $\rho^0 \to \pi^+\pi^-$ decay, weighted
with a factor $\sim E_\gamma^3$, where $E_\gamma$
is the photon energy in the $\eta$ center-of-mass system.
We generate $\tpi$ according to the distribution measured in~\cite{threepi}.
The simulation of $\eeg$ is analogous to $\pi^0 \to e^+e^-\gamma$
(``Dalitz decay'')~\cite{faessler}.

\begin{figure}
\includegraphics*[width=6.5in]{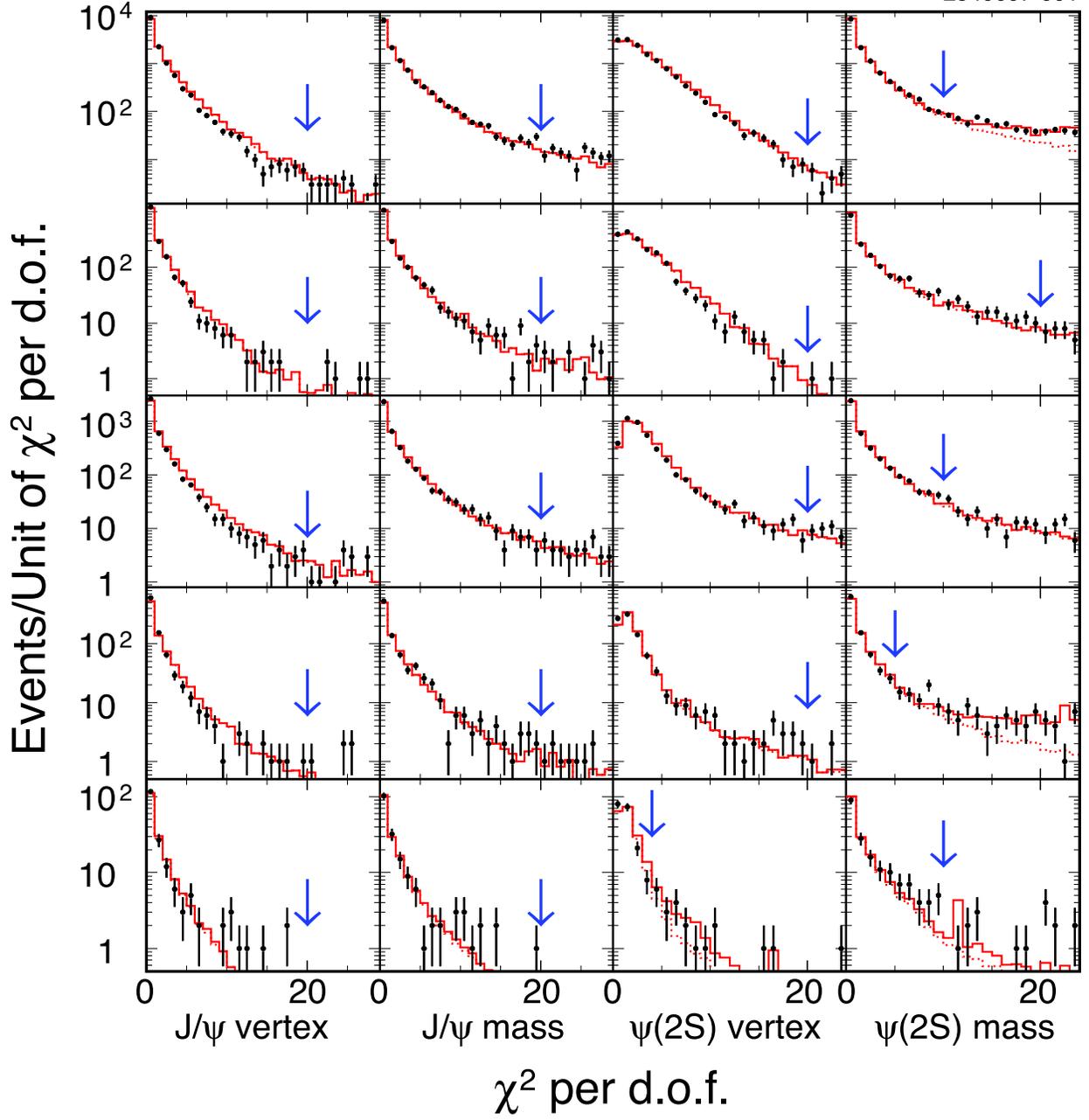}
\caption{Top to bottom: $\gga$, $\tpz$, $\tpi$,
$\ppg$, $\eeg$. For each channel, left to right: 
Goodness-of-fit for
$J/\psi$ vertex and mass fit, and for $\psi(2S)$
vertex and mass fit.
Points: data. Dotted line: Signal MC. Solid line: Sum of all MC. 
Arrows
indicate selection requirements. Cuts have been applied to
all quantities with the exception of the one plotted.  }
\label{fig:chisq}
\end{figure}

The event selection proceeds as follows.
We select the $J/\psi \to \ell^+\ell^-$ track candidates
within polar angles $|\cos \theta_{\ell^\pm}| < 0.83$,
adding bremsstrahlung photons within a
cone of $100\mrad$ around the track momentum vector
at the collision point. 
We identify leptons through the ratio of energy deposition
in the calorimeter associated with the track, $E$, 
to the track momentum measured in the drift chamber, $p$:
For electron (muon) candidates, we require $E/p$ values
of $> 0.85$ ($<0.25$) for one lepton and $>0.50$ ($<0.50$)
for the other. 
We impose kinematic constraints
by fitting the two lepton candidates to a common originating 
vertex (where the figure of merit is given by $\chijv$)
and to the $J/\psi$ mass ($\chijm$). We keep candidates that
have $\chijv < 20$ and $\chijm < 20$, which keeps signal
decays with high efficiency, as evident from Fig.~\ref{fig:chisq}.
The photons in the $\eta$ decay products are required to
be in the region of best calorimeter performance and least
material in front of the crystals, $|\cos \theta_\gamma| < 0.75$, 
and not be matched or close to a track's projection into
the calorimeter.

We then proceed to use kinematic constraints once more for improved
event cleanliness: 
The fitted $J/\psi$ and the $\eta$ decay products
are constrained, together with the beam spot~\cite{beamspot}, 
to a common vertex, and then to the $\psi(2S)$ mass.
This results in a very clean separation of final states.
We apply mode-dependent
restrictions on the quality of these fits,
denoted by $\chipv$ and $\chipm$, respectively.
Conversion events originating from $\eta \to \gamma\gamma$ decay
can fulfil the $\eeg$ pre-selection, but have
a poor $\psi(2S)$ vertex fit; hence
we apply a stricter cut of $\chipv < 4$  in
this channel (see Fig.~\ref{fig:chisq}). 
All other modes require $\chipv < 20$.
The mass fit has mode-dependent cuts, set as loosely
as possible while preserving sample cleanliness: 
$\chipm < 20$ for $\tpi$,
$< 10$ for $\gga$, $\tpz$, and $\eeg$,
$< 5$ for $\ppg$. 

\begin{figure}
\includegraphics*[width=6.5in]{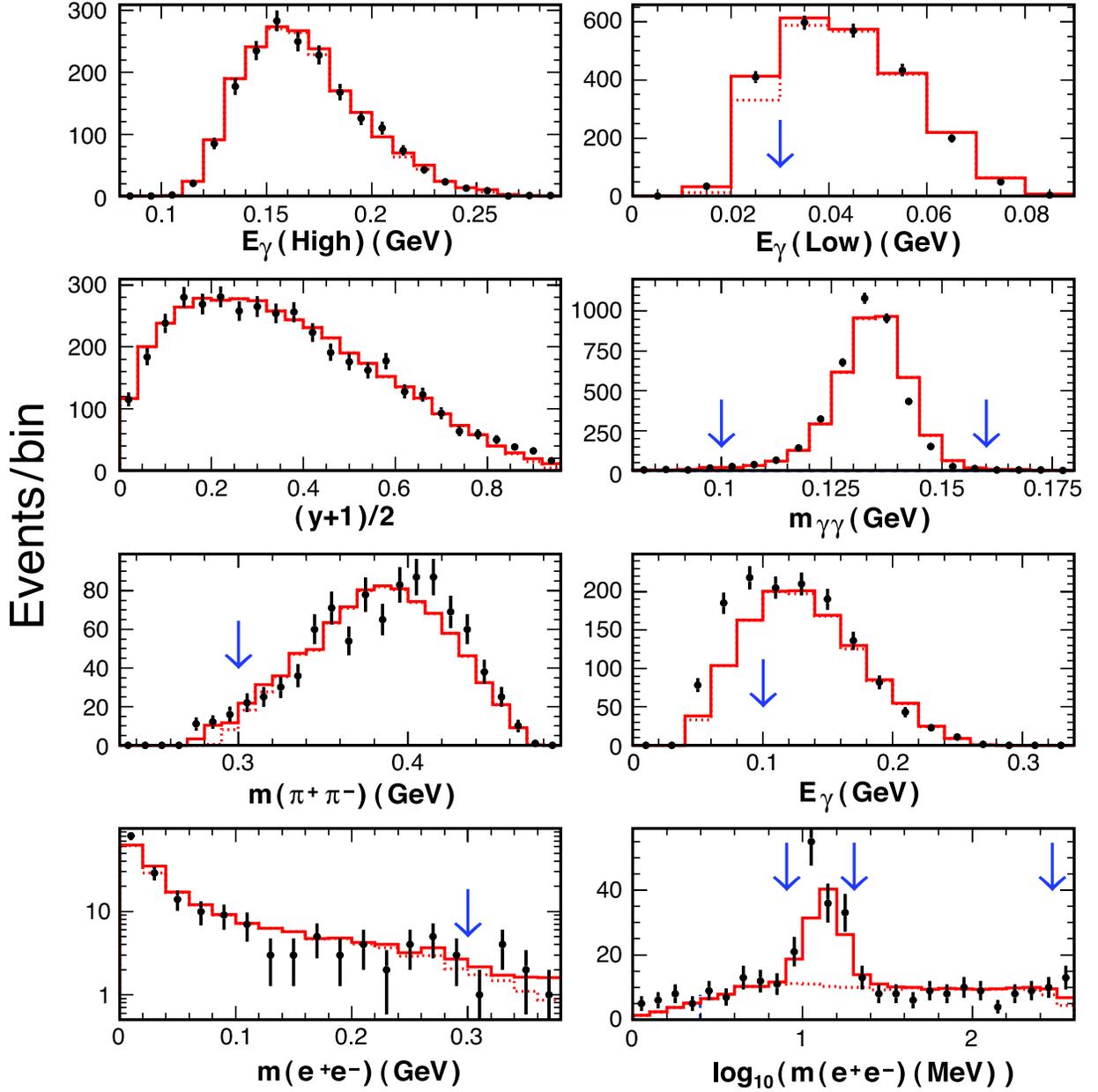}
\caption{Distributions for individual channels. Top row:
$\tpz$, highest and lowest photon energies.
Second row: $\tpi$, kinematic distribution of the three
pions, and two-photon invariant mass;
$y$ is a function
of the kinetic energy of the $\piz$ ($T_0$) and the
sum of the kinetic energies of all pions ($Q$): $y=(3T_0/Q)-1$.
Third row: $\ppg$, invariant mass of the two pions, and
photon energy. 
Fourth row: $\eeg$, invariant mass of the two electrons
on different horizontal and vertical scales.
Symbols as in Fig.~\ref{fig:chisq}.
}
\label{fig:permode}
\end{figure}

After this step, we define the following signal windows:
$p(J/\psi) = 170-230\mev/c$, and two ranges for $m(\eta)$:
$542-554\mev$  for $\ppg$,
$535-560\mev$  for all others.
Final state
specific characteristics are: (1) $\gga$: $E_\gamma> 200\mev$,
to suppress photons from $\psi(2S) \to \gamma\chi_{c1}$.
(2) $\tpi$: We search for two photons with $E_\gamma > 30\mev$
and $m(\gga) = 100-160\mev$, and 
constrain them to the $\piz$ mass. (3) $\tpz$: We
search for six photons, but do not attempt to make assignments
to $\piz$ candidates because doing so typically results in
multiple comparably probable assignments.
(4) $\ppg$: $E_\gamma > 100\mev$ and
$m(\pi^+\pi^-) > 300\mev$.
(5) $\eeg$: We add bremsstrahlung photons to the soft electrons
as with $J/\psi \to \ell^+\ell^-$, and the soft $e^\pm$
tracks must satisfy
$| \cos\theta| < 0.8$. In addition, we require
$m(e^+e^-) < 300\mev$. 
A substantial number of $\gga$ events with a conversion
survive the vertex restriction described above and fake
the $\eeg$ signature;
indeed, this type of background has necessitated substantial 
subtractions in previous measurements of this 
mode~\cite{eeg_cmd2,eeg_snd}. 
These conversions tend to occur at the discrete locations such as the
beam pipe and tracking chamber boundaries, but are
reconstructed as if they originated at the interaction
point. Consequently they create an artificial mass peak
near 10 MeV as seen in the lower right of 
Fig.~\ref{fig:permode}.
We remove the mass region $m(e^+e^-)=8-20\mev$ 
to suppress this background and the systematic uncertainties
associated with it.

For all five $\eta$ decay channels, 
we keep the two $J/\psi$ decay modes separate.
The fit quality for
data and simulation is compared in Fig.~\ref{fig:chisq}.
As a cross-check, we also perform the analysis {\sl without}
the $\psi(2S)$ kinematic fit: consistent results are obtained, 
but in most modes with far worse background conditions and
larger uncertainties.

The main backgrounds arise from cross-feed between the
$\eta$ modes and from other $\psi(2S) \to X J/\psi$
transitions, mostly $X=\pi^+\pi^-$, $\piz\piz$, 
and $\gamma\gamma$ through $\chi_{cJ}$. 
We select such exclusive event samples using selections
similar to the $\eta$ signal decays, including the kinematic
fits. Backgrounds from these $XJ/\psi$ channels 
into the $\eta$ signals are
then determined by scaling the MC predictions so as to match
the observed $XJ/\psi$ yields in data, and subtracted.
The statistical uncertainties of these subtractions are
accounted for. We find that $\gga$, $\tpz$, $\tpi$, and $\ppg$ 
have such backgrounds at the levels of 1-2\%. 
Examination of the $\eta$ mass sidebands revealed 
no discrepancy between data yields and MC estimate;
the only exception is $\ppg$, where data exceeds MC
by an amount which, when extrapolated into the
signal region, corresponds to a background of $(2.8 \pm 1.1)\%$
and is subtracted in addition to the other Monte Carlo
predictions. The mode $\eeg$ has a background of
about $5\%$ due mostly to $\gga$ conversions which survived the
$m(e^+e^-)$ and tight vertex fit restrictions.
Other (non-$J/\psi$) $\psi(2S)$ decays do not
fake the signal signature at any appreciable level.
We use a $20.7\invpb$ sample of data taken at a center-of-mass 
energy of $3.670\gev$ to estimate continuum background
(scaled by luminosity and energy dependence),
which is found to be negligible.

\begin{figure}
\includegraphics*[width=6.5in]{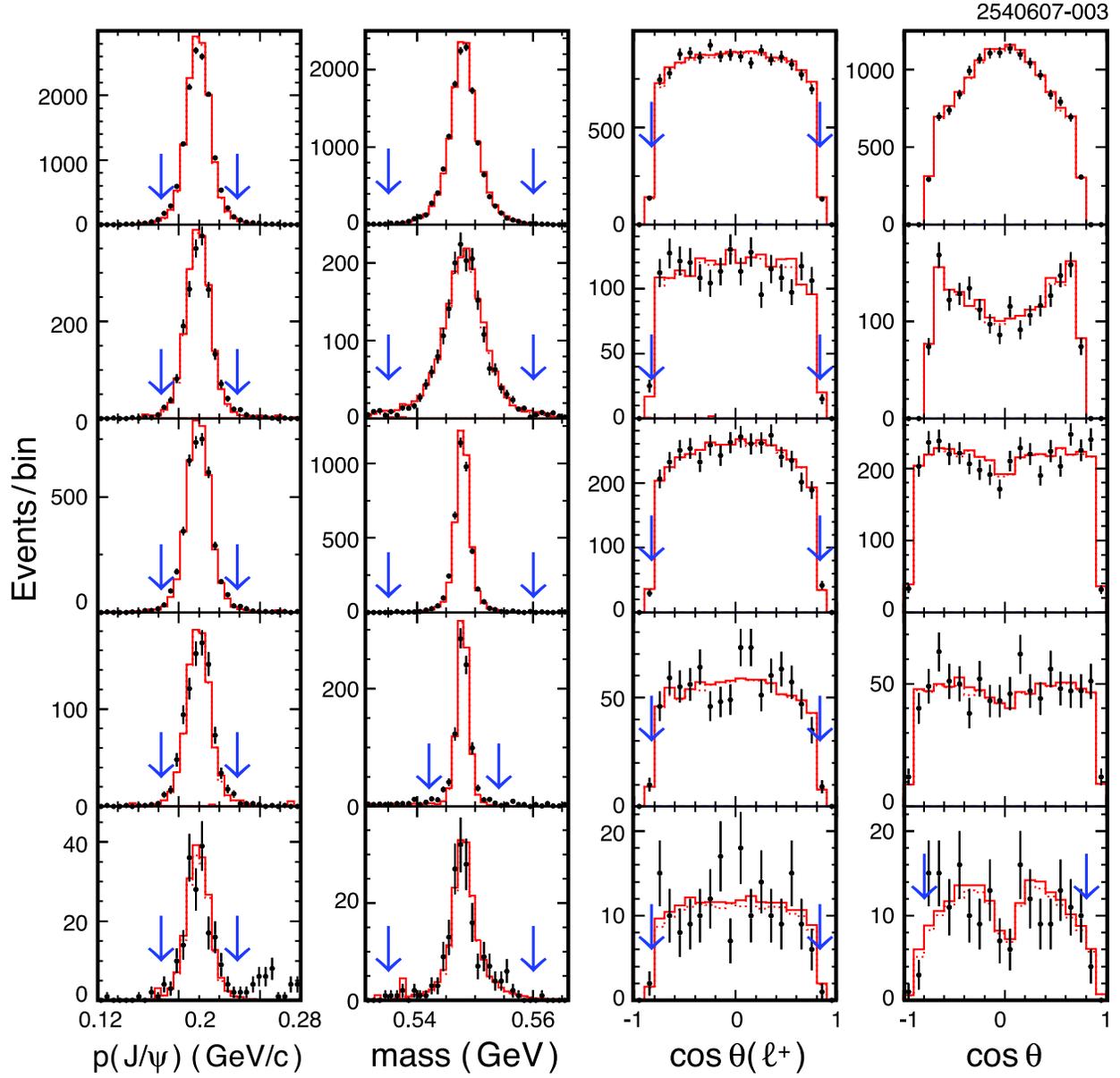}
\caption{Top to bottom: $\gga$, $\tpz$, $\tpi$,
$\ppg$, $\eeg$. For each channel, left to right:
$J/\psi$ momentum, $\eta$ mass, polar angle of
the positive lepton from the $J/\psi$ decay, and polar
angle of an $\eta$ decay product (most energetic shower
for $\gga$ and $\tpz$, positive track for $\tpi$,
$\ppg$, and $\eeg$).
Symbols as in Fig.~\ref{fig:chisq}.}
\label{fig:masses_and_thetas}
\end{figure}

All inspected experimental
observables show good agreement between data and 
the sum of our MC samples, normalized according to 
their relative population in the data. 
A selection of comparisons is shown in 
Figs.~\ref{fig:permode} and~\ref{fig:masses_and_thetas}.

Table~\ref{tab:yields} lists
observed yields and the estimated background.
We observe significant, clean, and unambiguous signals for all our target modes.

Our measurements are performed as ratios between efficiency-corrected
event yields of pairs of $\eta$ final states, separately 
for $J/\psi \to e^+e^-$ and $\mu^+\mu^-$.
This allows cancellation of all lepton-related systematic
uncertainties, such as track finding, lepton identification,
and $J/\psi$ fitting. We then proceed to combine the two
measurements of each ratio, where the $\eta$-related
uncertainties are treated as fully correlated. 
We note that the absolute detection efficiency for
$\psi(2S) \to \eta J/\psi$, $\eta \to \gga$, 
$J/\psi \to \ell^+\ell^-$ is about one third.

Sources of systematic uncertainty and the values
assigned are:
Track finding ($0.3\%$ per track, added linearly~\cite{xjpsi}), 
photon finding ($0.4\%$ per photon, added linearly~\cite{xjpsi}), 
sideband subtraction ($1.1\%$, $\ppg$ only),
trigger ($0.1$-$0.5\%$, mode-dependent), and
MC statistics ($0.4$-$1.0\%$, mode-dependent),
other effects in the detector simulation 
($0.5\%$).
We also make reasonable variations
in decay modeling at the MC generator level and
assign uncertainties accordingly:
$0.1\%$ for $\tpz$ to account for the slight deviation
from phase-space-prescribed decay observed 
in Ref.~\cite{threepizero},
$0.9\%$ for $\tpi$ based on the experimental uncertainty
of the slope parameter in Ref.~\cite{threepi},
$3\%$ for $\pi^+\pi^-\gamma$
to include a slightly different lineshape parameterization
of the intermediate $\rho^0$ meson,
and
$5\%$ for $e^+e^-\gamma$ to allow for changes
in the polar angle distribution of the $e^+$ from
the $\eta$ decay and in the $m(e^+e^-)$ spectrum
that remain consistent
with our measurements.
All uncertainties are
added in quadrature, except where correlations
between modes have to be observed.

\begin{figure}
\includegraphics*[width=6.5in]{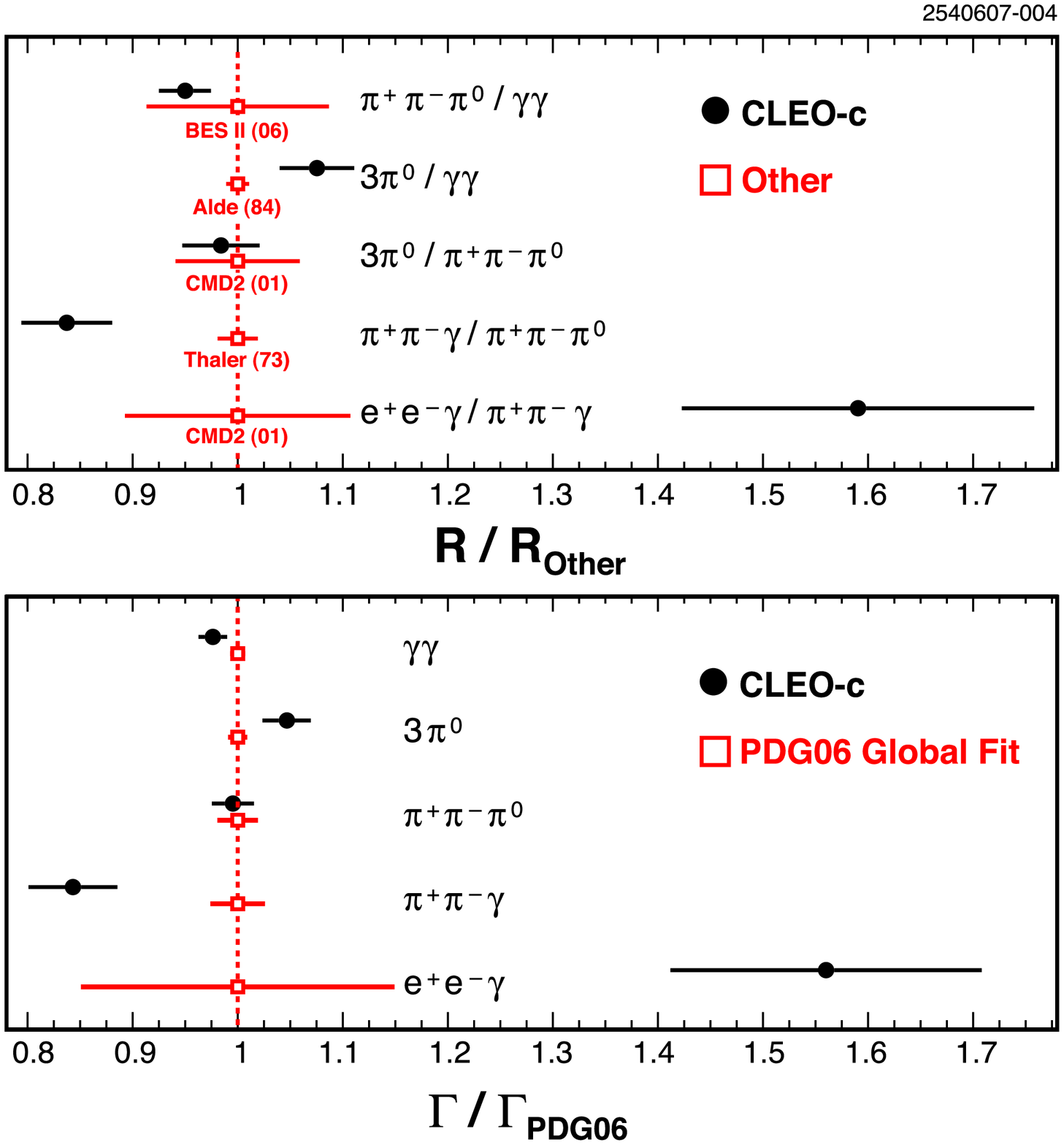}
\caption{Comparison of the results obtained in this
analysis with the most precise measurements from other
experiments~\cite{bestmeasts,Yao:2006px} (top), and 
the PDG 2006 global fits~\cite{Yao:2006px}.
}
\label{fig:comparison}
\end{figure}

The results for ratios of branching fractions
are shown in Table~\ref{tab:ratios}.
The $\chi^2$ for the ratios relative to $\gamma\gamma$ 
to agree between $J/\psi \to e^+e^-$ and $\mu^+\mu^-$ is
5.9 for four degrees of freedom, corresponding to a
confidence level of $\sim 20\%$.
We designate the following four ratios as constituting a
complete set, having minimal systematic correlation
with each other:
$\tpz/\gga$, $\tpi/\gga$, $\ppg/\tpi$, and $\eeg/\ppg$.
We compare to the single most precise other measurements
in Fig.~\ref{fig:comparison}.

Under the assumption that our five signal modes account for 
all of the $\eta$ decay modes, we combine the
ratios between them to form absolute branching fraction
measurements. 
Correlations between uncertainties are taken into account.
Other possible $\eta$ decay modes
are either forbidden and/or have been found to be
below 0.2\% in branching fraction ~\cite{Yao:2006px}: 
We include $0.3\%$ as a systematic uncertainty in 
the absolute branching fraction results. 
The results are presented in Table~\ref{tab:absolute},
together with those from PDG 2006~\cite{Yao:2006px} for
the global fit to all measurements.
In all five modes, the statistical uncertainty is larger
than or comparable to the systematic error.
A visual comparison can be found in Fig.~\ref{fig:comparison}.

To summarize, we have studied five $\eta$ decay
modes using the decay chain $\psi(2S) \to \eta J/\psi$,
$J/\psi \to ee$ and $\mu\mu$: $\eta \to \gga$,
$\tpz$, $\tpi$, $\ppg$, and $\eeg$. 
We have presented ratios between these modes
as well as absolute $\eta$ branching fractions to these final 
states.
This is the first analysis that covers this range
of $\eta$ decay modes, summing up to $99.9\%$ of
the known $\eta$ decays, and determines their absolute
branching fractions in the same experiment.
Several of the relative and derived absolute branching fractions
obtained in this analysis are either the most precise to date
or first measurements. In particular, we note that our result for
$\ppg$ is about $\sim 15\%$ ($3.2\sigma$) smaller than previous 
measurements, and for $\eeg$ is $\sim 57\%$ ($2.9\sigma$) larger.

We gratefully acknowledge the effort of the CESR staff
in providing us with excellent luminosity and running conditions.
D.~Cronin-Hennessy and A.~Ryd thank the A.P.~Sloan Foundation.
This work was supported by the National Science Foundation,
the U.S. Department of Energy, and
the Natural Sciences and Engineering Research Council of Canada.

\begin{table}
\caption{For each $\eta$ decay channel, the observed yields
in the $\psi(2S)$ on-resonance sample 
($N^{\psi(2S)}$),
background from cross-feed between $\eta$ modes
($N^{\mathrm{cf}}$), and background from other
$XJ/\psi$ decays ($N^{XJ/\psi}$)
separately for $J/\psi \to ee$ and $J/\psi \to \mu\mu$.
}
\begin{tabular}{c|r@{ $-$ }r@{ $-$ }r|r@{ $-$ }r@{ $-$ }r}
Channel 
 & \multicolumn{6}{c}{ $N^{\psi(2S)}  - N^{\mathrm{cf}} - N^{XJ/\psi} $} \\ 
 & \multicolumn{3}{c|}{$J/\psi \to e^+e^-$}  
 & \multicolumn{3}{c}{$J/\psi \to \mu^+\mu^-$} \\
 \hline 
$\gga$ &  6324 & 0 & 66 & 7376 & 0 & 114 \\ 
$\tpz$ &   850 & 0 & 18 & 1004 & 0 & 15 \\ 
$\tpi$ &  1884 & 4 & 12 & 2052 & 5 & 0 \\ 
$\ppg$ &   403 & 3 & 17 & 498 & 2 & 20 \\ 
$\eeg$ &    82 & 4 & 0 & 100 & 6 & 0 \\ 
\end{tabular} 
\label{tab:yields}
\end{table}

\begin{table}
\caption{Ratios of $\eta$ branching fractions.
For each combination, the efficiency ratio, separately for
$J/\psi \to e^+e^-$ and $J/\psi \to \mu^+\mu^-$, the level of
consistency between the $J/\psi \to e^+e^-$ and $\mu^+\mu^-$
result, expressed in units of Gaussian standard
deviations, $\sigma_{\mu\mu/ee}$, and the combined
result for the branching ratio. 
The dagger symbol indicates that this result is 
most precise measurement to date. 
}
\begin{tabular}{c|c|c|r|c@{ $\pm$ }c@{ $\pm$ }l}
Channel & \multicolumn{2}{c|}{eff. ratio} & $\sigma_{\mu\mu/ee}$& \multicolumn{3}{c}{branching fraction ratio} \\ & $\mu\mu$  & $ee$                &          &\multicolumn{3}{c}{\quad}    \\
 \hline 
$\tpz/\gga$ & $   0.15$ & $0.15$ & $  1.0$ &  0.884 & 0.022 & 0.019\\ 
$\tpi/\gga$ & $   0.50$ & $0.49$ & $ -2.2$ &  0.587 & 0.011 & 0.009$^\dagger $\\ 
$\ppg/\gga$ & $   0.63$ & $0.60$ & $  0.2$ &  0.103 & 0.004 & 0.004$^\dagger $\\ 
$\eeg/\gga$ & $   0.53$ & $0.52$ & $  0.1$ &  0.024 & 0.002 & 0.001$^\dagger $\\ 
$\tpz/\tpi$ & $   0.30$ & $0.32$ & $  2.1$ &  1.496 & 0.043 & 0.035$^\dagger $\\ 
$\ppg/\tpi$ & $   1.27$ & $1.24$ & $  1.1$ &  0.175 & 0.007 & 0.006 \\ 
$\eeg/\tpi$ & $   1.07$ & $1.06$ & $  0.5$ &  0.041 & 0.003 & 0.002$^\dagger $\\ 
$\eeg/\ppg$ & $   0.84$ & $0.86$ & $  0.0$ &  0.237 & 0.021 & 0.015 \\ 
\end{tabular} 
\label{tab:ratios}
\end{table}

\begin{table}
\caption{For each $\eta$ decay channel, absolute
branching fraction measurements for $J/\psi \to e^+e^-$
and $J/\psi \to \mu^+\mu^-$ combined, with statistical and
systematic uncertainties (middle column), as determined
in this work. The last column
shows the PDG fit result~\cite{Yao:2006px}. 
All but $\gamma\gamma$ are
first measurements~\cite{eegamma_comment}.
}
\begin{tabular}{c|r@{ $\pm$ }r@{ $\pm$ }l|r@{ $\pm$ }r}
 Channel & \multicolumn{3}{c|}{this work (\%)}  & \multicolumn{2}{c}{PDG~\cite{Yao:2006px} (\%)} \\ 
 \hline 
$\gga$ & $38.45 $ & $ 0.40 $ & $ 0.36$  & 39.38 & 0.26 \\ 
$\tpz$ & $34.03 $ & $ 0.56 $ & $ 0.49$  & 32.51 & 0.28 \\ 
$\tpi$ & $22.60 $ & $ 0.35 $ & $ 0.29$  & 22.7  \mbox{$\,$}& 0.4\mbox{$\ \,$} \\ 
$\ppg$ & $3.96 $ & $ 0.14 $ & $ 0.14$  &  4.69 & 0.11 \\ 
$\eeg$ & $0.94 $ & $ 0.07 $ & $ 0.05$  &  0.60 & 0.08 \\ 
\end{tabular} 
\label{tab:absolute}
\end{table}



\begin{thebibliography}{99}
\bibitem{beamspot} 
The beam spot is the center of the ellipsoidal  $e^+e^-$ collision region,
and is determined from finding
a common vertex position in hadronic
events that occur in contiguous 30-60 minute
periods of stable beam conditions.
\end{thebibliography}
\end{document}